\documentclass[%twocolumn,
showpacs,preprintnumbers,amsmath,amssymb]{revtex4}
\usepackage{amsmath}

\begin{document}

\tolerance=5000

\def\be{\begin{equation}}
\def\ee{\end{equation}}
\def\bea{\begin{eqnarray}}
\def\eea{\end{eqnarray}}
\def\tr{{\rm tr}\, }
\def\nn{\nonumber \\}
\def\e{{\rm e}}

\title{Reconstructing the universe history, from inflation to
acceleration, \\ with phantom and canonical scalar fields}

\author{Emilio Elizalde$^{1}$,  Shin'ichi Nojiri$^{2}$,
Sergei D. Odintsov$^{3}$\footnote{Also at Lab. Fundam. Study, Tomsk State
Pedagogical University, Tomsk}, Diego S\'{a}ez-G\'{o}mez$^{1}$, and
Valerio Faraoni$^{4}$\medskip }

\affiliation{$^{1}$Consejo Superior de Investigaciones Cient\'\i ficas
ICE/CSIC-IEEC \\ Campus UAB, Facultat de Ci\`encies, Torre
C5-Parell-2a pl, E-08193 Bellaterra (Barcelona) Spain}

\affiliation{$^{2}$Department of Physics, Nagoya University, Nagoya
464-8602, Japan}

\affiliation{$^{3}$Instituci\`{o} Catalana de Recerca i Estudis Avan\c{c}ats
(ICREA) and Institut de Ci\`{e}ncies de l'Espai (IEEC-CSIC), Campus UAB,
Facultat de Ci\`encies, Torre C5-Parell-2a pl, E-08193 Bellaterra
(Barcelona) Spain}

\affiliation{$^{4}$Physics Department, Bishop's University,
Sherbrooke, Qu\'{e}bec, Canada J1M 0C8}

\begin{abstract}

We consider the reconstruction technique in theories with
a single or multiple (phantom and/or canonical) scalar fields.
With the help of several examples, it is demonstrated explicitly that the
universe expansion history, unifying early-time inflation
and late-time acceleration, can be realized in scalar-tensor
gravity.
This is generalized to the theory of a scalar field coupled
non-minimally to the curvature and to a Brans-Dicke-like theory.
Different examples of unification of inflation with cosmic
acceleration, in which de Sitter, phantom, and quintessence type fields
play the fundamental role---in different combinations---are
worked out. Specifically, the
frame dependence and stability properties of de Sitter
space scalar field theory are studied. Finally, for two-scalar
theories, the late-time acceleration and early-time inflation epochs are
successfully reconstructed, in realistic situations in which
the more and more stringent observational
bounds are satisfied, using the freedom of choice of
the scalar field potential, and of the kinetic factor.
 \end{abstract}


\maketitle

\section{Introduction}

The increasing amount and precision of observational data
demand that theoretical cosmological models be as
realistic as possible in their description of the evolution of
our universe. The discovery of late-time cosmic acceleration
brought into this playground
a good number of dark energy models (for a recent review, see \cite{DDE}) which
aim at describing the observed accelerated expansion, which seems
to have started quite recently on the redshift scale. Keeping in
mind the possibility, which is well compatible with the
observational data, that the effective equation of state
parameter $w$ be less than $-1$, phantom
cosmological models \cite{phantom2,phantom,LTC,phantom3} share a
place in the list of theories capable to explain dark
energy. Ideal fluid and scalar field
quintessence/phantom models still remain among the easiest and most
popular constructions. Nevertheless, when working with these
models, one should
bear in mind that such theories are at best effective descriptions of
the early/late universe, owing to a number of well-known problems.

Even in such a situation, scalar field models still remain quite
popular candidates for dark energy. An additional problem
with these theories--which traditionally has not been
discussed in depth--is that a good
mathematical theory must not be limited to the description of
a single side of the cosmic evolution:  it
should  rather provide a unified description of the whole
expansion history of the universe,
from the inflationary epoch to the onset of cosmic acceleration,
and beyond. Note that a similar drawback is also typical of
inflationary models, most
of which have problems with ending inflation and also fail
to describe realistically the late-time universe.

The purpose of this work is to show that, given a certain scale
factor (or Hubble parameter) for the universe expansion
history, one can in fact reconstruct it from a specific scalar
field theory. Using multiple scalars, the
reconstruction becomes easier due to the extra
freedom brought by the arbitrariness in the
scalar field potentials and kinetic factors. However, there are
subtleties in these cases that can be used advantageously, and
this makes the study of those models even more interesting.

Specifically, in this work we overview the reconstruction
technique for scalar theories with one, two, and an arbitrary
number, $n$, of fields. After that, many explicit examples are
presented in which a unified, continuous  description of the
inflationary era and of the late-time cosmic acceleration epoch
is obtained in a rather simple and natural way.

The organization of this paper is as follows: in Sec.~II we
consider a universe filled with matter and show that it is
possible to obtain both inflation and accelerated expansion at
late times by using a single scalar field. Realistic
examples are worked out in order to illustrate this fact.
Sec.~III is devoted to the theory of a scalar coupled
non-minimally to gravity through the Ricci curvature. Late-time
acceleration is explicitly discussed in this model, using two
specific examples. In Sec.~IV we discuss the issue of
reconstruction for a non-minimally coupled scalar field theory,
including Brans-Dicke theory, again clarifying this case with
the help of an example. Sec.~V deals with de Sitter space in
scalar-tensor theory and studies issues relevant for the
conformal transformation to the Einstein frame, arriving to
stability conditions that are important in the study of
the future evolution of the late-time accelerated era. In
Sec.~V, we discuss the case of several scalar fields, beginning
with the case of two scalar fields minimally coupled to gravity
(such models have been used, {\em e.g.}, in reheating scenarios
after inflation). We study an explicit example and then the
general case of $n$ scalar fields.
Sec.~VII addresses our final goal, namely the
reconstruction of inflation and cosmic acceleration from a
scalar field theory, by means of a two-scalar model that
reproduces cosmological constraints in each epoch.
Further, the cosmic acceleration is reproduced with a pair of
scalar fields plus an ordinary matter term, where the observed
cosmological density parameter
($\Omega _{\rm DE}\simeq 0.7$) and the equation of state (EoS)
parameter ($w_{\rm DE}\simeq -1$) are actually reproduced.
Finally, Sec.~VIII contains the conclusions.

\section{Unified inflation and late time acceleration in scalar
theory}

Let us consider a universe filled with matter with equation
of state $p_{m}=w_{m}\rho_{m}$ (here $w_{m}$ is a
constant) and a scalar field
which only depends on time. We will show that it is possible to
obtain both inflation and accelerated expansion at late times
by using a single scalar field $\phi$ (see also \cite{UNPHC} and
\cite{UPIAC}).
In this case, the action is
\begin{equation}
S=\int dx^{4}\sqrt{-g}\left[ \frac{1}{2\kappa^{2}}R
 - \frac{1}{2} \omega (\phi)
\partial_{\mu} \phi \partial^{\mu }\phi -V(\phi )+L_{m}\right]\ ,
\label{eq:1}
\end{equation}
where $ \kappa^2 =8\pi G$, $V(\phi )$ being the scalar potential
and $ \omega (\phi)$ the kinetic function, respectively,
while $L_{m}$ is the matter
Lagrangian density.
Note that for convinience the kinetic factor is introduced.
At the final step of calculations, scalar field maybe always redefined so
that kinetic factor is absorbed in its definition.
 As we work in a spatially
flat Friedmann-Robertson-Walker (FRW) spacetime, the metric is
given by
\begin{equation}
ds^{2}=-dt^{2}+a^2(t) \sum_{i=1}^{3} dx_{i}^{2}\ .
\label{eq:1.2}
\end{equation}
The corresponding FRW equations are written as
\be
H^{2} = \frac{\kappa ^{2}}{3}\left( \rho _{m}
+\rho_{\phi}\right)\ , \quad \quad \dot H = -\frac{\kappa ^{2}}{2}\left(
\rho _{m}+p_{m}+\rho _{\phi }+p_{\phi }\right)\ ,
\label{eq:1.3}
\ee
with $\rho_{\phi}$ and $p_{\phi}$ given by
\be
\rho _{\phi } = \frac{1}{2} \omega (\phi )\, {\dot \phi}^{2}
+V(\phi)\ ,\quad \quad
p_{\phi } = \frac{1}{2} \omega (\phi ) \, {\dot \phi}^{2}-V(\phi)\ .
\label{eq:1.4}
\ee
Combining Eqs.~(\ref{eq:1.3}) and (\ref{eq:1.4}), one obtains
\be
\omega (\phi ) \, \dot{\phi ^{2}}
= -\frac{2}{\kappa^{2}}\dot{H}-(\rho _{m}+p_{m})\ ,\quad \quad
V(\phi ) = \frac{1}{\kappa ^{2}}
\left( {3H}^{2}+\dot{H} \right) -\frac{\rho_{m}-p_{m}}{2}\ .
\label{eq:1.5}
\ee
As the matter is not coupled to the scalar field, by using
energy conservation one has
\be
\dot{\rho _{m}}+3H(\rho _{m}+p_{m})=0\ ,\quad
\dot{\rho _{\phi }} +3H(\rho _{\phi }+p_{\phi })=0\ .
\label{eq:1.6}
\ee
 From the first equation, we get $\rho_{m}=\rho _{m0}a^{-3(1+ w_{m})}$.
We now consider the theory in which $V(\phi)$ and $\omega(\phi)$ are
\be
\omega (\phi ) = -\frac{2}{\kappa ^{2}}f^{\prime }(\phi )
 -(w_{m}+1)F_{0} \e^{-3(1+w_{m})F(\phi )}\ ,\quad
V(\phi ) = \frac{1}{\kappa ^{2}}
\left[ {3f(\phi)}^{2}+f^{\prime }(\phi ) \right]
+\frac{w_{m}-1}{2}F_{0}\, \e^{-3(1+w_{m})F(\phi )}\ ,
\label{eq:1.6a}
\ee
where $f(\phi) \equiv F'(\phi )$, $F$ is an arbitrary (but
twice differentiable) function of $\phi$, and $F_{0}$ is an
integration constant. Then, the following solution is found
(see
\cite{OCDE, UNPHC,UPIAC, MtDE}):
\be
\label{ST1}
\phi =t\ , \quad H(t)=f(t)\ ,
\ee
which leads to
\begin{equation}
a(t)=a_{0}\e^{F(t)}, \qquad a_{0}=\left(
\frac{\rho _{m0}}{F_{0}}\right) ^{\frac{1}{3(1+w_{m})}}.
\label{eq:1.7}
\end{equation}
We can study this system by analyzing the effective
EoS parameter which, using the FRW
equations, is defined as
\begin{equation}
w_{\rm eff} \equiv \frac{p}{\rho }=-1-\frac{2\dot{H}}{{3H}^{2}}
\; ,
\label{eq:1.8}
\end{equation}
where
\be
\rho = \rho _{m}+\rho _{\phi }\ ,\quad
p = p_{m}+p_{\phi }\ . \label{1.8b}
\ee
Using the formulation above, one can present  explicit
examples of reconstruction as follows.

\subsection{Example 1}

As a first example, we consider the following model:
\begin{equation}
f(\phi )=h_{0}^{2}\left( \frac{1}{t_{0}^{2}-\phi ^{2}}
+ \frac{1}{\phi^{2}+t_{1}^{2}}\right)\ . \label{eq:1.9}
\end{equation}
Using the solution~(\ref{eq:1.7}), the Hubble parameter and the
scale factor are given by
\be
H=h_{0}^{2}\left(
\frac{1}{t_{0}^{2}-t^{2}}+\frac{1}{t^{2}+t_{1}^{2}}\right)\ ,\quad
a(t)=a_{0}\left( \frac{t+t_{0}}{t_{0}-t}\right)
^{\frac{h_{0}^{2}}{2t_{0}}
}e^{\frac{h_{0}^{2}}{t_{1}}\arctan\frac{t}{t_{1}}}\ .
\label{eq:1.9a}
\ee
As one can see, the scale factor vanishes at $t=-t_{0}$, so we
can fix that
point as corresponding to the creation of the universe. On the
other hand, the kinetic function and the scalar
potential are given by Eqs.~(\ref{eq:1.6a}),
hence
\bea
\omega (\phi ) &=& -\frac{8}{\kappa^{2}}
\frac{h_{0}^{2}(t_{1}^{2}+t_{0}^{2})
\left( \phi ^{2}-\frac{t_{1}^{2}+t_{0}^{2}}{2}\right)
\phi}{(t_{1}^{2}
+\phi^{2})^{2}(t_{0}^{2}-\phi ^{2})^{2}}-(w_{m}+1)F_{0}
\e^{-3(w_{m}+1)F(\phi )}\ ,\nn
&&\nonumber \\
V(\phi ) &=& \frac{h_{0}^{2}(t_{1}^{2}+t_{0}^{2})}
{\kappa^{2}(t_{1}^{2}+\phi^{2})^{2}
(t_{0}^{2}-\phi ^{2})^{2}}\left[ 3h_{0}^{2}(t_{1}^{2}+t_{0}^{2})+4
\phi \left( \phi ^{2}-\frac{t_{1}^{2}+t_{0}^{2}}{2}\right) \right]
+\frac{w_{m}-1}{2}F_{0}\ e^{-3(w_{m}+1)F(\phi )}\ ,
\eea
where $F_{0}$ is an integration constant and
\be
\label{ST2}
F(\phi )=\frac{h_{0}^{2}}{2t_{0}}\ln\left( \frac{\phi
+t_{0}}{t_{0}-\phi }\right) +\frac{h_{0}^{2}}{t_{1}}\arctan \frac{\phi }{t_{1}}\ .
\ee
Then, using Eq.~(\ref{eq:1.8}), the effective EoS
parameter is written as
\begin{equation}
w_{\rm eff}=-1-\frac{8}{3h_{0}^{2}}\frac{t(t-t_{+})(t+t_{-})}{
(t_{1}^{2}+t_{0}^{2})^{2}} \;, \label{eq:1.10}
\end{equation}
where $t_{\pm }=\pm \sqrt{\frac{t_{0}^{2}-t_{1}^{2}}{2}}$. There
are two phantom phases that occur when $ t_{-}<t<0$ and
$t>t_{+}$, and
another two non-phantom phases for $-t_{0}<t<t_{-}$ and
$0<t<t_{+}$, during which $
w_{\rm eff}>-1$ (matter/radiation-dominated epochs). The first
phantom phase can be interpreted as an inflationary epoch, and
the second one as corresponding to the current accelerated
expansion, which will end in a Big Rip singularity when $
t=t_{0}$. Note that superacceleration ({\em i.e.}, $\dot{H}>0$)
is due to the negative sign of the kinetic
function $\omega (\phi)$, as for ``ordinary'' phantom fields
(to which one could reduce by redefining the scalar $\phi$).

\subsection{Example 2}

As a second example, we consider the choice
\begin{equation}
f(\phi)=\frac{H_0}{t_s-\phi}+\frac{H_1}{\phi^2}\ .
\label{1.17}
\end{equation}
We take $H_0$ and $H_1$ to be  constants and $t_s$ as the Rip
time, as specified
below. Using (\ref{eq:1.6a}), we find that the kinetic
function and the scalar potential are
\bea
\omega(\phi) &=& -\frac{2}{\kappa^2}\left[
\frac{H_0}{(t_s-\phi)^2}-\frac{2H_1}{\phi^2}
\right] -(w_m+1)F_0\left(
t_s-\phi\right)^{3(1+w_m)H_0}\exp\left[
\frac{3(1+ w_m)H_1}{\phi}\right]\ , \nn
&&\nonumber \\
V(\phi) &=& \frac{1}{\kappa^2}\left[
\frac{H_0(3H_0+1)}{(t_s-\phi)^2}
+\frac{H_1}{\phi^3}\left(\frac{H_1}{\phi}-2 \right)\right]
+\frac{w_m-1}{2}
F_0\left(t_s-\phi \right)^{3(1+ w_m)H_0}
\e^{\frac{3(1+ w_m)H_1}{\phi}}\ ,
\label{1.18}
\eea
respectively. Then, through the solution~(\ref{eq:1.7}), we
obtain the Hubble parameter and the scale factor
\be
H(t)=\frac{H_0}{t_s-t}+\frac{H_1}{t^2}\ ,\quad
a(t)=a_0\left( t_s-t\right)^{-H_0} \e^{-\frac{H_1}{t}}\ .
\label{1.19}
\ee
Since $a(t)\to 0^{+} $ for $t\to 0$, we can fix $t=0$ as the
beginning of the universe. On the other hand, at $t=t_s$ the
universe reaches a
Big Rip singularity, thus we keep $t<t_s$. In order to study the different stages that our
model will pass through, we calculate the acceleration parameter and the first
derivative of the Hubble parameter. They are
\be
\dot H=\frac{H_0}{(t_s-t)^2}-\frac{2H_1}{t^3}\ ,\quad
\frac{\ddot a}{a}=H^2+\overset{.}{H}=\frac{H_0}{(t_s-t)^2}(H_0+1)
+ \frac{H_1}{t^2}\left(\frac{H_1}{t^2}-\frac{2H_1}{t}
+\frac{2H_0}{t_s-t}
\right)\ .
\label{1.20}
\ee
As we can observe, for $t$ close to zero, $\ddot a/a>0$, so that
the universe is accelerated during some time. Although this is
not a phantom
epoch, since $\dot H<0$, such stage can be interpreted as
corresponding to the
beginning of inflation. For $t>1/2$ but $t\ll t_s$, the universe
is in a
decelerated epoch ($\ddot a/a<0$). Finally, for $t$ close to $t_s$,
it turns out that $\dot H>0$, and then the universe is
superaccelerated, such acceleration being of phantom nature and
ending in a Big Rip singularity at $t=t_s$.

\subsection{Example 3}

Our third example also exhibits unified inflation and
late time
acceleration, but in this case we avoid phantom phases and, therefore, Big Rip
singularities. We consider the following model:
\begin{equation}
f(\phi )=H_{0}+\frac{H_{1}}{\phi ^{n}}\ , \label{1.11}
\end{equation}
where $H_{0}$ and $H_{1}>0$ are constants and $n$ is a positive
integer (also constant). The case $n=1$ yields an initially
decelerated universe and a late time acceleration phase. We
concentrate on cases corresponding to $n>1$ which gives, in
general, three epochs: one of early acceleration (interpreted as
inflation), a second decelerated phase and,
finally, accelerated expansion at late times. In this model, the
scalar potential and the kinetic parameter are given, upon
use of Eqs.~(\ref{eq:1.6a}) and (\ref{1.11}), by
\bea
\omega (\phi ) &=&\frac{2}{\kappa ^{2}}\frac{nH_{1}}
{\phi^{n+1}}
 -(w_{m}+1)F_{0} \, \e^{-3(w_{m}+1)\left( H_{0}\phi
 -\frac{H_{1}}{(n-1)
\phi^{n-1}}\right) }, \label{1.12} \\
&&\nonumber\\
V(\phi ) &=&\frac{1}{\kappa ^{2}}\frac{3}{\phi ^{n+1}}\left[
\frac{\left(
H_{0}\phi ^{n/2}+H_{1}\right)^{2}}{\phi ^{n-1}}-\frac{nH_{1}}{3}\right]
+\frac{w_{m}-1}{2}F_{0} \, \e^{-3(w_{m}+1)\left( H_{0}\phi -
\frac{ H_{1}}{(n-1)\phi ^{n-1}}\right) }.
\eea
Then, the Hubble parameter given by the solution (\ref{eq:1.7})
can be written as
\be
H(t) = H_{0}+\frac{H_{1}}{t^{n}}\ , \label{1.13} \quad
a(t) = a_{0}\exp \left[
H_{0}t -\frac{H_{1}}{(n-1)t^{n-1}}\right]\ .
\ee
We can fix $t=0$ as the beginning of the universe because at
this point $a\to 0$, so $t>0$. The effective EoS
parameter~(\ref{eq:1.8}) is
\begin{equation}
w_{\rm eff}=-1+\frac{2nH_{1}t^{n-1}}{\left(
H_{0}t^{n}+H_{1}\right) ^{2}}\ .
\label{1.14}
\end{equation}
Thus, when $t\to 0$ then $ w_{\rm eff}\to -1$ and we
have an acceleration epoch, while for $t\to \infty $,
$w_{\rm eff} \to -1$ which
can be interpreted as late time
acceleration. To find the phases of acceleration and deceleration for $t>0$,
we study $\ddot a/a$, given by:
\begin{equation}
\frac{\ddot a}{a}=\dot H +H^{2}=-\frac{nH_{1}}{t^{n+1}}
+\left( H_{0}+\frac{H_{1}}{t^{n}}\right) ^{2}\ . \label{1.15}
\end{equation}
For sufficiently large values of $n$
we can find two positive zeros of this function, which means
two corresponding
phase transitions. They happen, approximately, at
\begin{equation}
t_{\pm }\approx \left[ \sqrt{nH_{1}} \,\, \frac{\left( 1\pm
\sqrt{1-\frac{4H_{0} }{n}} \, \right)}{2H_{0}} \, \right]^{2/n}\ ,
\label{1.16}
\end{equation}
so that, for $0<t<t_{-}$, the universe is in an accelerated
phase interpreted as an inflationary epoch; for $t_{-}<t<t_{+}$
it is in a decelerated phase (matter/radiation dominated); and,
finally, for $t>t_{+}$ one obtains late time acceleration, which
is in agreement with the current cosmic expansion.

We now consider how the exit from inflation could be realized.
First, we should note that if one refines the scalar field as
$\varphi = \int^\phi d\phi \sqrt{\omega(\phi)}$ in the non-phantom phase or
$\varphi = \int^\phi d\phi \sqrt{-\omega(\phi)}$ in the phantom phase,
then the action (1) %(\ref{eq:1})
has the following form
\be
\label{k1B}
S=\int d^4 x \sqrt{-g}\left\{\frac{1}{2\kappa^2}R \mp \frac{1}{2}\partial_\mu
\varphi
\partial^\mu \varphi  - \tilde V(\varphi)\right\} + S_m\, .
\ee
Here $\tilde V(\varphi)\equiv V\left(\phi(\varphi)\right)$.
The minus sign corresponds to the non-phantom phase and the plus sign to
the phantom phase.
For example, in the model (21) %(\ref{1.11})
when $\phi$ is small (and therefore in the early universe), we
find $\varphi = - 2\sqrt{2n H_1}/(n-1)\kappa\phi^{(n-1)/2}$ and
$\tilde V(\varphi) \propto \varphi^{4 + 4/(n-1)}$. We should note that in the
early universe ($\phi\to 0$),
$\varphi$ is large. Then, the potential $\tilde V(\varphi)$ is of the slow roll
type. When the value of $\varphi$ is large, the energy of the vacuum
is large, which generates inflation. The value of $\varphi$ slowly becomes
small and the energy of the vacuum, and with it the curvature, decrease.
When the curvature becomes small enough inflation could stop.
For the successful exit from inflation we may need one more scalar field,
as in the hybrid inflation scenario \cite{linde}. In the model (21), %(\ref{1.11})
by construction, inflation ends as a purely classical theory
when $\ddot a = a (H^2 + \dot H )=0$, that is, for $t=t_e$ satisfying
\be
\label{inf2}
0=\left(H_0 + \frac{H_1}{t_e^n}\right)^2 - \frac{nH_1}{t_e^{n+1}}\ .
\ee
If we include, however, quantum effects the value of the scalar field
$\varphi$ jumps to a larger value---as a consequence of the quantum
fluctuations---and the vacuum acquires higher energy, what generates inflation
again. In order to suppress the probability of those effects, in the hybrid
inflation model at least one more scalar field and its coupling with $\varphi$
had to be introduced.

\subsection{Example 4}

As our last example, we consider another model unifying
early universe inflation and the accelerating expansion
of the present universe. We may choose $f(\phi)$ as
\be
\label{uf0}
f(\phi)=\frac{H_i + H_l c  \e^{2\alpha\phi}}{1 + c
\e^{2\alpha\phi}}\ ,
\ee
which gives the Hubble parameter
\be
\label{uf1}
H(t)=\frac{H_i + H_l c \e^{2\alpha t}}{1 + c \e^{2\alpha t}}\ .
\ee
Here $H_i$, $H_l$, $c$, and $\alpha$ are positive constants.
In the early universe ($t\to -\infty$), we find that $H$
becomes a constant
$H\to H_i$ and at late times ($t\to +\infty$), $H$ becomes a
constant again $H\to H_l$.
Then $H_i$ could be regarded as the effective cosmological
constant driving inflation, while $H_l$ could be a small
effective constant generating the late acceleration.
Then, we should assume $H_i \gg H_l$.
Hence, if we consider the model with action
\bea
\label{uf2}
S &=& \int d^4 x \sqrt{-g}\left\{\frac{R}{2\kappa^2} -
\frac{\omega(\phi)}{2}\partial_\mu \phi
\partial^\mu \phi - V(\phi)\right\}\ ,\nn
&&\nonumber \\
\omega(\phi) &\equiv& - \frac{f'(\phi)}{\kappa^2}
= \frac{2\alpha\left(H_i - H_l\right)c
\e^{2\alpha\phi}}{\kappa^2
\left(1 + c\e^{2\alpha\phi}\right)^2}\ ,\nn
&&\nonumber \\
V(\phi) &\equiv& \frac{3f(\phi)^2 + f'(\phi)}{\kappa^2}
= \frac{3 H_i^2 + \left\{6 H_i H_l - 2\alpha\left( H_i - H_l
\right)\right\}c\e^{2\alpha\phi}
+ c^2 H_l^2  \e^{4\alpha\phi}}{\kappa^2 \left(1 +
c\e^{2\alpha\phi}\right)^2}\ ,
\eea
we can realize the Hubble rate given by (\ref{uf1}) with $\phi=t$.
If we redefine the scalar field as
\be
\label{uf2b}
\varphi \equiv \int d\phi \sqrt{\omega(\phi)}
= \frac{\e^{\alpha\phi}}{\kappa}\sqrt{\frac{2(a-b)c}{\alpha}}\ ,
\ee
the action $S$ in (\ref{uf2}) can be rewritten in the canonical
form
\be
\label{uf2c}
S =  \int d^4 x \sqrt{-g}\left\{\frac{R}{2\kappa^2} -
\frac{1}{2}\partial_\mu \varphi
\partial^\mu \varphi - \tilde V(\varphi)\right\}\ ,
\end{equation}
where
\be
\tilde V(\varphi) = V(\phi)
= \frac{3 H_i^2 + \frac{\kappa^2 \alpha \left\{6 H_i H_l
 - 2 \alpha\left( H_i - H_l \right)\right\}}{2\left( H_i - H_l
\right)}\varphi^2
+ \frac{3  \kappa^4 H_l^2 \alpha^2}{4\left(H_i -
H_l\right)^2}\varphi^4}
{\kappa^2  \left(1 + \frac{\kappa^2 \alpha}{2\left( H_i - H_l
\right)}\varphi^2\right)^2}\; .
\ee
One should note that $\phi\to - \infty$ corresponds to
$ \varphi\to 0$ and $V\sim 3H_i^2/\kappa^2$, while  $\phi\to +
\infty$  corresponds to $\varphi\to \infty$ and $V\sim
3H_l^2/\kappa^2$, as expected.
At early times ($\varphi\to 0$), $\tilde V(\varphi)$ behaves  as
\be
\label{uf3c}
\tilde V(\varphi) \sim \frac{3H_i^2}{\kappa^2} \left\{
1 - \frac{\kappa^2 \alpha \left(3 H_i + \alpha
\right)}{3H_i^2}\varphi^2
+ {\cal O}\left(\varphi^2\right)\right\}\ .
\ee
At early times, $\phi<0$ and therefore, from Eq.~(\ref{uf2b}),
we find
$\kappa\varphi \sqrt{\alpha/2\left(H_i - H_l\right)c}
\ll 1$, from which it follows that
\bea
\label{uf3d}
&& \frac{1}{3\kappa^2}\frac{{\tilde V}'(\varphi)}{\tilde
V(\varphi)^2}
\sim \frac{4\alpha^2 \kappa^2\left(3 H_i +
\alpha\right)^2}{27H_i^4}\varphi^2
< \frac{8\alpha \left(3 H_i + \alpha\right)^2 \left(H_i -
H_l\right)c}{27H_i^4}\ ,\nn
&&\nonumber \\
&& \frac{1}{3\kappa^2}\frac{\left|{\tilde V}''(\varphi)
\right|}{\tilde V(\varphi)}
\sim \frac{2\left(3H_i + \alpha\right)\alpha}{9H_i^2}\ .
\eea
Then, if $\alpha\ll H_i$, the slow-roll conditions can be
satisfied.

We may include matter with constant EoS parameter $w_m$. Then
$\omega(\phi)$ and $V(\phi)$
are modified as
\bea
\label{uf3}
\omega(\phi) &\equiv& - \frac{f'(\phi)}{\kappa^2}
 - \frac{w_m + 1}{2}g_0 \e^{-3(1+w_m)g(\phi)}\ ,\nn
&&\nonumber \\
V(\phi) &\equiv& \frac{3f(\phi)^2 + f'(\phi)}{\kappa^2}
+ \frac{w_m -1}{2}g_0 \e^{-3(1+w_m)g(\phi)}\ ,\nn
&&\nonumber \\
g(\phi) &\equiv& \int d\phi f(\phi) = H_l \phi + \frac{H_i -
H_l}{2}
\ln \left(c + \e^{-2\alpha \phi}\right)\ .
\eea
The matter energy density is then given by
\be
\label{uf4}
\rho_m = \rho_0 a^{-3(1+w_m)} = g_0 \e^{-3(1+w_m)g(t)}
= g_0 \left( c + \e^{-2\alpha t}\right)^{-3(1+w_m)(H_i-H_l)/2} \e^{-3(1+w_m)H_l t}\ .
\ee
In the early universe $t\to -\infty$, $\rho_m$ behaves as
\be
\label{uf5}
\rho_m \sim g_0 \e^{3(1+w_m)\left(2\alpha - H_l\right) t}\ .
\ee
On the other hand, the energy density of the scalar field behaves as
\be
\label{uf6}
\rho_\varphi = \frac{1}{2}{\dot\varphi}^2 + \tilde V(\varphi)
= \frac{\omega(\phi)}{2}{\dot\phi}^2 + V(\phi) \to \frac{3H_i^2}{\kappa^2}\ .
\ee
Then, if $2\alpha < H_l$ (and $w_m>-1$), the matter contribution
could be neglected in comparison with the
scalar field contribution.

Now let the present time be $t=t_0$. Then, we find that
\bea
\label{uf7}
\Omega_m &\equiv& \frac{\kappa^2 \rho_m}{3 H^2} \nn
&&\nonumber \\
&=& \frac{\kappa^2 g_0 \e^{4\alpha t_0}
\left( c + \e^{-2\alpha t_0}\right)^{-3(1+w_m)(H_i-H_l)/2\alpha + 2} \e^{-3(1+w_m)H_l t_0}}
{3\left( H_i + H_l c \e^{2\alpha t_0}\right)^2}\;,
\eea
and $\Omega_\phi = 1 - \Omega_m$. If we assume $\alpha t_0 \gg
1$, we find
\be
\label{uf8}
\Omega_m \sim \frac{\kappa^2 g_0}{3H_l^2}
c^{-3(1+w_m)(H_i-H_l)/2\alpha} \e^{-3(1+w_m)H_l t_0} \ .
\ee
Hence, we may choose the parameters so that $\Omega_m \sim
0.27$, which could be consistent with the observed data.
This model provides a quite realistic picture of the unification
of the inflation with the present cosmic speed-up.

\section{Accelerated expansion in the non-minimally
curvature-coupled scalar theory}

In the preceding section we have considered an action,
(\ref{eq:1}), in which the
scalar field is minimally coupled to gravity. In the present
section, the scalar
field couples to gravity through the Ricci scalar
(see \cite{think} for a review on cosmological
applications). We begin from the action
\begin{equation}
S=\int d^{4}x\sqrt{-g} \left[(1+f(\phi))\frac{R}
{\kappa^{2}}-\frac{1}{2} \omega (\phi ) \partial_{\mu }\phi
\partial^{\mu }\phi -V(\phi )\right]\ ,
\label{2.1}
\end{equation}
where $f(\phi )$ is an arbitrary function of the scalar field
$\phi $. Then,
the effective gravitational coupling depends on $\phi $, as
$\kappa_{eff}=\kappa [1+f(\phi)]^{-1/2}$. One can work in the
Einstein frame,
by performing the scale transformation
\begin{equation}
g_{\mu \nu }=[1+f(\phi )]^{-1} \widetilde{g}_{\mu \nu }\ .
\label{2.2}
\end{equation}
The tilde over $g$ denotes an Einstein
frame quantity. Thus, the action~(\ref{2.1}) in such a frame
assumes the form \cite{CT}
\begin{equation}
S=\int d^{4}x\sqrt{-\widetilde{g}}\left\{
\frac{\widetilde{R}}{2 \kappa ^{2}}
 -\left[ \frac{ \omega (\phi )}{2(1+f(\phi))}+\frac{6}
{\kappa^{2}(1+f(\phi ))}
\left( \frac{ d(1+f(\phi)^{1/2})}{d\phi }\right) ^{2}\right]
\partial_{\mu }\phi
\partial^{\mu } \phi -\frac{V(\phi)}{[ 1+f(\phi )]^{2}}\right\}\
.
\label{2.3}
\end{equation}
The kinetic function can be written as $W(\phi )=\frac{ \omega
(\phi )}{1+f(\phi)}
+\frac{3}{\kappa ^{2}(1+f(\phi ))^{2}}
\left( \frac{df(\phi )}{d\phi }\right) ^{2}$, and the extra term in the
scalar potential can be absorbed by defining the new potential
$U(\phi )=\frac{ V(\phi )}{ \left[ 1+f(\phi ) \right]^2}$, so
that we recover
the action~(\ref{eq:1}) in the
Einstein frame, namely
\begin{equation}
S=\int dx^{4} \sqrt{-\widetilde{g}}\left(
\frac{\widetilde{R}}{ \kappa ^{2}}-
\frac{1}{2} \, W( \phi ) \, \partial _{\mu }\phi \partial ^{\mu}
\phi -U(\phi)\right)\ .
\label{2.4}
\end{equation}
We assume that the metric is FRW and spatially flat in this
frame
\begin{equation}
d\widetilde{s}^{2}=-d\widetilde{t}^{2}
+\widetilde{a}^{2}(\widetilde{t})\sum_{i}dx_{i}^{2}\ ,
\label{2.5a}
\end{equation}
then, the equations of motion in this frame are given by
\begin{eqnarray}
&& \widetilde{H}^{2}=\frac{\kappa ^{2}}{6}\rho _{\phi } \;,\\
&&\nonumber \\
&& \dot{\widetilde{H}}=-\frac{\kappa ^{2}}{{4}}\left( \rho
_{\phi } + p_{\phi}\right)\ , \label{2.6}\\
&&\nonumber \\
&& \frac{d^2 \phi}{d\tilde{t}^2} +3\tilde{H} \,
\frac{d\phi}{d\tilde{t}} +\frac{1}{2W(\phi)}\left[ W'(\phi)
\left( \frac{d\phi}{d\tilde{t}}\right)^2 +2U'(\phi) \right] =0
\;,
\end{eqnarray}
where $\rho _{\phi }=\frac{1}{2}W(\phi ){\dot \phi}^{2} +
U(\phi )$, $p_{\phi }=\frac{1}{2}W(\phi ){\dot \phi}^{2} -
U(\phi )$, and
the Hubble parameter is $\widetilde{H}
\equiv \frac{1}{\widetilde{a}}
\frac{d\widetilde{a}}{d\widetilde{t}}$. Then,
\be
W(\phi )\dot{\phi}^{2} =-4\dot{\widetilde{H}}\ ,\quad \quad
U(\phi )=6\widetilde{H}^{2}+2\dot{\widetilde{H}}. \label{2.7}
\ee
Note that $ \dot{\widetilde{H}} >0$ is equivalent to $W<0$;
superacceleration is due to the ``wrong'' (negative) sign of the
kinetic energy, which is the distinctive feature of a phantom
field. The scalar field could be redefined to eliminate
the factor $W(\phi)$, but this would not correct the sign of the
kinetic energy.

If we choose $W(\phi)$ and $U(\phi)$ as $ \omega(\phi)$ and
$V(\phi)$ in~(\ref{eq:1.6a}),
\be
W (\phi ) = -\frac{2}{\kappa ^{2}}g^{\prime }(\phi )\ ,\quad
U(\phi ) = \frac{1}{\kappa ^{2}} \left[ {3 g(\phi )}^{2}+ g'
(\phi ) \right]\ ,
\label{ST4}
\ee
by using a function $g(\phi)$ instead of $f(\phi)$
in~(\ref{eq:1.6a}), we find a solution as in
(\ref{ST1}),
\begin{equation}
\phi =\widetilde{t}\ , \quad \widetilde{H}(\widetilde{t})=g(\widetilde{t})\ .
\label{2.8}
\end{equation}
In (\ref{ST4}) and hereafter in this section,  we have dropped
the matter contribution for simplicity.

We consider the de Sitter solution in this frame,
\begin{equation}
\widetilde{H}=\widetilde{H}_{0}=\mbox{const.\ } \to \
\widetilde{a}( \widetilde{t})
=\widetilde{a}_{0}e^{\widetilde{H}_{0}\widetilde{t}_{{}}} \;.
\label{2.9}
\end{equation}
We will see below that accelerated expansion can be obtained
in the original
frame corresponding to the Einstein frame~(\ref{2.4}) with
the solution~(\ref{2.9}),
by choosing an appropriate function $f(\phi )$. From (\ref{2.9})
and
the definition
of $W(\phi )$ and $U(\phi )$, we have
\be
W(\phi )=0 \ \to \ \omega (\phi )=-\frac{3}{\left[ 1+f(\phi)
\right] \kappa^{2}}
\left[ \frac{df(\phi )}{d\phi }\right]^{2}\ ,\quad
U(\phi )=\frac{6}{\kappa ^{2}}\widetilde{H}_{0}^{2}\ \to \
V(\phi )=\frac{6}{\kappa ^{2}}\widetilde{H}_{0}^{2}[1+f(\phi )]^{2}\ . \label{2.10}
\ee
Thus, the scalar field has a non-canonical kinetic term in
the original frame, while in the Einstein frame the latter can
be
positive, depending on $W(\phi)$. The correspondence between
conformal frames can be made explicit through the conformal
transformation~(\ref{2.2}). Assuming a spatially flat FRW metric
in the
original frame,
\begin{equation}
ds^{2}=-dt^{2}+a^{2}(t)\sum_{i=1}^3 dx_{i}^{2}\ , \label{2.10a}
\end{equation}
then, the relation between the time coordinate and the scale
parameter in these frames is given by
\be
t=\int \frac{d\widetilde{t}}{([1+f(\widetilde{t})]^{1/2}}\ ,
\quad a(t)=[1+f(\widetilde{t})]^{-1/2} \,
\widetilde{a}(\widetilde{t})\ .
\label{2.11}
\ee
Now let us discuss the late-time acceleration in the model under
discussion.
%%%%%%%%%%%%%%%%%%%

\subsection{Example 1}

As a first example, we consider the coupling function
between the scalar field and the Ricci scalar
\begin{equation}
f(\phi )=\frac{1-\alpha \phi }{\alpha \phi }\ , \label{2.12}
\end{equation}
where $\alpha $ is a constant. Then, from~(\ref{2.10}), the
kinetic function $ \omega(\phi)$ and the
potential $V(\phi)$ are
\be
\omega(\phi)=-\frac{3}{\kappa^{2}\alpha^{2}}
\, \frac{1}{\phi^{3}}\ ,
\quad \quad
V(\phi)=\frac{6\widetilde{H}_{0}}{\kappa^{2}
\alpha^{2}}\frac{1}{\phi^{2}}\ ,
\label{2.13}
\ee
respectively. The solution for the current example is found to
be
\be
\phi (t)=\widetilde{t} =\frac{1}{\alpha }\left(
\frac{3\alpha}{2} \, t\right)^{2/3}\ , \quad \quad
a(t)=\widetilde{a}_{0}\left( \frac{3\alpha }{2} \,
t\right)^{1/3}
\exp \left[ \frac{\widetilde{H}_{0}^{{}}}{\alpha } \left(
\frac{3\alpha}{2} \, t\right)^{2/3}\right]\ .
\label{2.14}
\ee
We now calculate the acceleration parameter to study the
behavior of the scalar parameter in the original frame,
\begin{equation}
\frac{\ddot a}{a}=-\frac{2}{9}\frac{1}{t^{2}}+\widetilde{H}_{0}
\left( \frac{2}{3\alpha }\right)^{1/3}\left[ \frac{1}{t^{4/3}}
+ \widetilde{H}_{0}\left( \frac{2}{3\alpha }\right)^{1/3}\frac{1}{t^{2/3}}\right]\ .
\label{2.15}
\end{equation}
We observe that for small values of $t$ the acceleration is
negative; after that we get accelerated expansion for large
$t$; finally, the universe ends with zero acceleration as $t\to
\infty $. Thus, late time accelerated expansion is reproduced by
the action~(\ref{2.1}) with the function $f(\phi )$ given by
Eq.~(\ref{2.12}).

\subsection{Example 2}

As a second example, consider the function
\begin{equation}
f(\phi )=\phi -t_{0}\ .
\label{2.16}
\end{equation}
 From (\ref{2.11}), the kinetic term and the scalar potential are, in this case,
\be
\omega(\phi)=-\frac{3}{\kappa ^{2}}\frac{1}{(1+\phi-t_{0})}\ ,
\quad \quad
V(\phi)=\frac{6\widetilde{H}_{0}^{2}}{\kappa ^{2}}(1+\phi-t_{0})\ .
\label{2.17}
\ee
The solution in the original (Jordan) frame reads
\be
\phi(t)=\frac{t^{2}}{4}+t_{0}-1\ , \quad
a(t)= \frac{2\widetilde{a}_{0}}{t} \exp\left[\widetilde{H}_{0}
\left(\frac{t^{2}}{4} +t_{0}-1\right) \right]\ ,
\ee
and the corresponding acceleration is
\begin{equation}
\frac{\ddot a}{a}=\frac{1}{t^{2}}\left[\frac{\widetilde{H}_{0}t^{2}}{2}
+\left(\frac{\widetilde{H}_{0}t^{2}}{2}-1 \right)^{2}+1 \right]\ .
\end{equation}
Notice that this solution describes acceleration at every time $t$ and, for
$t\to \infty$, the acceleration tends to a constant value, as in de Sitter spacetime,
hence similar to what happens in the Einstein frame.
Thus, we have proved here that it is possible to reproduce accelerated expansion in both frames,
by choosing a convenient function for the coupling $f(\phi)$.

\section{Reconstruction of non-minimally coupled scalar field
theory}

We now consider the reconstruction problem in the non-minimally
coupled scalar field theory, or the Brans-Dicke theory. We begin
with the same scalar-tensor theory with constant
parameters $\phi_0$ and $V_0$:
\be
\label{BD1}
S=\int d^4 x \sqrt{-g} \left[ \frac{R}{2\kappa^2}
 - \frac{1}{2}\partial_\mu \phi \partial^\mu \phi
 - V(\phi) \right] \ ,\quad \quad V(\phi)=V_0 \,
\e^{-2\phi/\phi_0}\ ,
\ee
which admits the exact solution
\be
\label{BD2}
\phi =\phi_0\ln \left|\frac{t}{t_1}\right|\ ,\quad
H=\frac{\kappa^2\phi_0^2}{2 t}\ , \quad
t_1^2 \equiv \frac{\gamma \phi_0^2 \left(\frac{3\gamma
\kappa^2\phi_0^2}{2} - 1\right)}{2V_0}\ .
\ee
We choose $\phi_0^2 \kappa^2>2/3$ and $V_0>0$ so that
$t_1^2>0$. For this solution, the metric is given by
\be
\label{BD3}
ds^2 = - dt^2 + a_0^2
\left(\frac{t}{t_0}\right)^{\kappa^2\phi_0^2}\sum_{i=1}^3
\left(dx^i\right)^2\ ,
\ee
which can be transformed into the conformal form:
\be
\label{BD4}
ds^2 = a_0^2 \left(\frac{\tau}{ \tau_0}
\right)^{-\frac{\kappa^2 \phi_0^2}{\frac{\kappa^2\phi_0^2}{2} - 1}}
\left( - d\tau^2 + \sum_{i=1}^3 \left(dx^i\right)^2\right)\ .
\ee
Here
\be
\label{BD5}
\frac{\tau}{\tau_0} =
 - \left(\frac{t}{t_0}\right)^{-\frac{\kappa^2\phi_0^2}{2} +1 }\ ,\quad
\tau_0\equiv \frac{t_0}{a_0 \left(\frac{\kappa^2\phi_0^2}{2} - 1\right)}\ .
\ee
and therefore, by using (\ref{BD2}), one finds
\be
\label{BD6}
\frac{\tau}{\tau_0}=
 - \left(\frac{t_1}{t_0}\right)^{-\frac{\kappa^2}{2} +1}
\e^{- \frac{1}{-\frac{\kappa^2}{2} +1} \frac{\phi}{\phi_0}}\ .
\ee

We now consider an arbitrary cosmology given by the metric
\be
\label{BD7}
d{\tilde s}^2 = f(\tau)\left( - d\tau^2
+ \sum_{i=1}^3 \left(dx^i\right)^2\right)\ ,
\ee
where $\tau$ is the conformal time. Since
\be
\label{BD8}
ds^2=\frac{1}{f(\tau)}a_0^2 \left(\frac{\tau}{\tau_0}
\right)^{-\frac{\kappa^2\phi_0^2}{\frac{\kappa^2\phi_0^2}{2} - 1}}
d\tilde s^2 = \e^{\varphi} d\tilde s^2\ , \quad
\e^{\varphi} \equiv \frac{a_0^2
\left(\frac{t_1}{t_0}\right)^{\kappa^2\phi_0^2} \e^{\kappa^2 \phi_0 \phi}}
{f\left( - \tau_0
\left(\frac{t_1}{t_0}\right)^{-\frac{\kappa^2}{2} +1}
\e^{- \frac{1}{-\frac{\kappa^2}{2} +1}
\frac{\phi}{\phi_0}}\right)}\ ,
\ee
if we begin with the action in which $ g_{\mu\nu}$ in
(\ref{BD1}) is replaced by $\e^{\varphi}{\tilde g}_{\mu\nu}$,
\be
\label{BD9}
S= \int d^4 x \sqrt{-\tilde g} \, \e^{\varphi(\phi) }\left\{
\frac{R}{2\kappa^2} - \frac{1}{2}\left[
1 - 3\left(\frac{d\varphi}{d\phi}\right)^2\right]
\partial_\mu \phi \partial^\mu \phi
 - \e^{\varphi(\phi)} V(\phi) \right\}\ ,
\ee
we obtain the solution~(\ref{BD7}).

\subsection{Example}

By using the conformal time $\tau$, the metric of de Sitter
space
\be
\label{dS}
ds^2 = - dt^2 + \e^{2H_0 t}\sum_{i=1}^3 \left(dx^i\right)^2\ ,
\ee
can be rewritten as
\be
\label{dS2}
ds^2 = \frac{1}{H_0^2\tau^2}\left( - d\tau^2 +
\sum_{i=1}^3 \left(dx^i\right)^2\right)\ .
\ee
Here $\tau$ is related to $t$ by $\e^{- H_0 t}=- H_0\tau$.
Then $t\to -\infty$ corresponds to $\tau \to +\infty$ and $t\to
+\infty$ corresponds to $\tau \to 0$.

As an example of $f(\tau)$ in (\ref{BD7}), we may consider
\be
\label{dS3}
f(\tau)=\frac{\left(1 + H_L^2\tau^2\right)}{H_L^2\tau^2\left(1
+ H_I^2 \tau^2\right)}\ ,
\ee
where $H_L$ and $H_I$ are constants. At early times in the
history of the
universe $\tau\to \infty$ (corresponding to $t\to -\infty$),
$f(\tau)$ behaves as
\be
\label{dS4}
f(\tau)\to \frac{1}{H_I^2\tau^2}\ .
\ee
Then the Hubble rate is given by a constant $H_I$, and therefore
the universe is asymptotically de Sitter space, corresponding to
inflation. On the other hand, at late times $\tau\to 0$
(corresponding to $t\to + \infty$), $f(\tau)$ behaves as
\be
\label{dS5}
f(\tau)\to \frac{1}{H_L^2\tau^2}\ .
\ee
Then the Hubble rate is again a constant $H_L$, which may
correspond to the late time acceleration of the universe.
This proves that our reconstruction program can be applied directly to the
non-minimally coupled scalar theory.

\section{de Sitter space in scalar-tensor theory}

When studying scalar field cosmology in a spatially flat
FRW universe from the dynamical systems
point of view, it is often convenient to redefine the scalar
field $\phi$ used in the previous sections in such a way that
its kinetic energy density has a canonical form (apart from the
sign). Instead of the field $\phi$ appearing in Eq.~(\ref{2.4}),
one can use
\be
\label{extra1}
\sigma\equiv \int d\phi \, \, \sqrt{ \left| W(\phi) \right|} \;,
\ee
in terms of which the action~(\ref{2.4}) becomes
\be
S=\int d^4 x \sqrt{-g} \, \left[
\frac{R}{2\kappa^2}-\frac{\epsilon}{2}\, \partial^{\mu}\sigma
\partial_{\mu}\sigma -\bar{U}( \sigma) \right] \;,
\ee
where $\epsilon=\mbox{sign}(W)$ and
$\bar{U}(\sigma) =U\left[ \phi(\sigma) \right]$ (this
redefinition, however, can not change the sign of the kinetic
energy of the phantom field to make it positive - for this
purpose one would have to make the scalar field purely
imaginary through a sort of Wick rotation).

In discussions of the phase space of spatially flat scalar field
cosmology, the Hubble parameter $H$ and the scalar field
$\sigma$ (or $\phi$) constitute a natural choice of dynamical
variables. The phase space is a two-dimensional curved surface
embedded in a three-dimensional space and this suffices to
guarantee the absence of chaos in the dynamics \cite{VMS}.
Moreover, the only fixed points of the dynamical system
are de Sitter spaces with constant scalar field $\left( H_0,
\sigma_0 \right)$. For the solution described by
Eqs.~(\ref{2.8}) and (\ref{2.9}),
the scalar field redefinition~(\ref{extra1}) yields
$\sigma=$const.$\equiv \sigma_0$ and $ \left( \tilde{H}_0,
\sigma_0 \right)$ is a
fixed point of the system. The fixed point nature of a
particular solution does depend on the specific choice of
dynamical variables: for example, the solution~(\ref{2.8}) and
(\ref{2.9}) is a fixed point with the choice $\left( H, \sigma
\right)$ but not with the choice $\left( H, \phi \right)$. While
the solution is the same, it is convenient to study the
dynamics using $\sigma$ instead of $\phi$. This is particularly
important for detailed calculations of the stability of de
Sitter space using gauge-invariant variables (see below), as
these calculations greatly simplify in a de Sitter background.

One may wonder whether the fixed point (or even the attractor)
nature of de Sitter spaces is lost when performing the
conformal transformation to the Einstein frame used in the
previous sections in order to find exact solutions. This is not
the case, but a little care is needed because, in general,
acceleration of the universe $\ddot{a}>0$ in the Jordan frame
does not imply cosmic acceleration $\frac{d^2 \tilde{a}}{
d\tilde{t}^2} >0$ in the Einstein frame, and
vice-versa since
\be
\frac{ d^2 \tilde{a} }{d\tilde{t}^2}=\Omega^{-1} \left[ \ddot{a}
+\frac{ \dot{\Omega} }{\Omega}\,\dot{a}+ \frac{\left( \Omega
\ddot{\Omega} - \dot{\Omega}^2 \right) }{\Omega^2} \,
a\right]
\ee
(here an overdot denotes differentiation with respect to the
Jordan frame comoving time $t$). However, a de Sitter space in
the Jordan
frame is mapped into a de Sitter space in the Einstein frame (and
vice-versa). A general conformal transformation of the metric
$ g_{\mu\nu}\rightarrow \tilde{g}_{\mu\nu}=\Omega^2 \,g_{\mu\nu}$,
where the conformal factor $\Omega $ depends on
the scalar field present in the theory (for example, as in
Eq.~(\ref{2.2})), yields the rescaled FRW line element
\be
d\tilde{s}^2=\Omega^2 ds^2=-d\tilde{t}^2
+\tilde{a}^2\left(\tilde{t}\right) \left( dx^2+dy^2+dz^2 \right)
\ee
with $d\tilde{t}=\Omega \, dt$ and $\tilde{a}=\Omega \, a$.
Therefore, the relation between the Hubble parameters of the two
conformal frames is
\be
\tilde{H}\equiv \frac{1}{\tilde{a}} \,
\frac{d\tilde{a}}{d\tilde{t}}=\frac{1}{\Omega} \left(
H+\frac{\dot{\Omega}}{\Omega} \right)=
\frac{1}{\Omega} \left(
H+\frac{d\Omega}{d\tilde{t} } \right) \;.
\ee
Since $\Omega=\Omega(\sigma)$,
it is clear that a Jordan frame fixed point $\left( \dot{H},
\dot{\sigma} \right) =\left( 0,0 \right)$ is mapped into an
Einstein frame fixed point $\left( \dot{\tilde{H}},
\dot{\sigma} \right) =\left( 0,0 \right)$.
Moreover, a small perturbation of this fixed point
in the Jordan frame corresponds
to a small perturbation in the Einstein frame, and stability of
the Jordan frame fixed point corresponds to stability of the
corresponding Einstein frame stationary point. In fact, assume
that
\begin{eqnarray}
H (t) &= & H_0+\delta H (t) \;, \\
&&\nonumber \\
\sigma (t) &=& \sigma_0+\delta \sigma (t) \;,
\end{eqnarray}
in the Jordan frame, where $\left| \delta H(t) /H_0 \right|$ and
$\left| \delta \sigma (t) /\sigma_0 \right|$ are small
perturbations which, for simplicity, are taken here to be
homogeneous. Then, in the Einstein frame, $ \tilde{H}=
\tilde{H}_0+\delta \tilde{H} $ with
\be
\delta \tilde{H}=\frac{1}{\Omega_0}\left[ \delta H +\frac{
\Omega_0 '
}{\Omega_0}\left( \delta \dot{\sigma} -H_0 \delta \sigma \right)
\right]
\ee
(where a prime denotes differentiation with respect to
$\sigma$ and the zero subscript denotes quantities evaluated
in
the background de Sitter space). The Einstein frame perturbation
stays small if the Jordan frame
perturbations are small, hence an attractor in the Jordan frame
corresponds to an Einstein frame attractor (the converse is not
always true, see \cite{Abreuetal} for a counterexample). This
property is
crucial when using conformal transformations to study slow-roll
inflation which describes dynamics around a de Sitter attractor.
It is the presence of a de Sitter attractor that justifies the
use of the slow-roll approximation $H(t)=H_0+\delta H(t)$ in
inflation, and
the fact that the presence and attractor nature of de Sitter
space (subject to certain conditions) are
guaranteed also in the Einstein frame ultimately justifies the
use of conformal techniques in the study of slow-roll
inflation. To
summarize, in general, cosmic acceleration in one conformal
frame does not correspond to acceleration in the conformally
related frame(see the related discussion in \cite{capo}),
however slow-roll
inflation in the Jordan
frame corresponds to slow-roll inflation in the Einstein frame.
This is relevant also for late time de Sitter-like expansion in
a universe dominated by quintessence.

The previous discussion is restricted to homogeneous
(space-independent) perturbations of de Sitter space, but it
can be extended to more general (space-dependent)
{\em inhomogeneous} perturbations. The latter are
more problematic because of the notorious gauge-dependence
problems associated with them, and they are best described
using gauge-independent methods. A linear stability condition
of de Sitter space against inhomogeneous perturbations was
derived in~\cite{myperts}. This
is a
necessary condition for de Sitter space to be an attractor in
the $\left( H, \sigma \right)$ phase space, however it is not
sufficient because it only ensures stability to first order in
the gauge-invariant variables (stability to higher, or to all
orders, can usually be established only numerically and with
the restriction to homogeneous perturbations, see, {\em e.g.},
\cite{CarvalhoSaa}).

For a theory described by the action
\be
S=\int d^4 x \sqrt{-g} \left[ \varphi\left( \phi, R \right)
 -\frac{\omega(\phi )}{2} \,\partial^{\mu}\phi\partial_{\mu} \phi
 -V(\phi) \right] \;,
\ee
the gauge-invariant linear stability condition is \cite{myperts}
\be
\frac{
\frac{\partial^2\varphi}{\partial \phi^2} \left. \right|_0
 -\frac{d^2V}{d\phi^2} \left. \right|_0 +
\frac{R_0 \varphi_{\phi\, R}^2}{F_0} }
{\omega_0 \left( 1+ 3\frac{\varphi_{\phi\, R}^2}{\omega_0 F_0}
\right) } \leq 0 \;,
\ee
where $ F\equiv \frac{\partial \varphi}{\partial R}$ and
$\varphi_{\phi \, R }\equiv \frac{\partial^2 \varphi}{\partial R
\partial \phi} $. For the action~(\ref{2.1}), this
condition becomes
\be
\frac{
R_0 \left[ f_0'' \left( 1+f_0 \right)+ \left(
f_0'\right)^2 \right] -2\kappa^2 V_0''}{
2\kappa^2 \omega_0 \left( 1+f_0 \right) +3\left( f_0'
\right)^2} \leq 0 \;,
\label{stability}
\ee
where $f_0 \equiv f(\phi_0), f_0'\equiv f'(\phi_0)$, {\em etc.}
and assuming that $1+f(\phi_0) >0$ in order to guarantee a
positive
effective gravitational coupling. For an ``ordinary'' phantom
field in general relativity it is $\varphi \left( \phi, R
\right)=R$, $\omega=-1$, and the condition~(\ref{stability})
reduces to $V_0''\leq 0$, {\em i.e.}, the de Sitter space is
stable if the potential has a maximum to which the phantom field
can climb and settle in during the dynamical evolution. This
behaviour, which is opposite to that of an ordinary scalar
field, is due to the negative sign of the kinetic energy of the
phantom \cite{phantom}.

The FRW equations give two conditions for the existence of de
Sitter fixed points:
\begin{eqnarray}
&& R_0 \, \frac{\partial \varphi}{\partial R}
\left.\right|_0=2\varphi_0-V_0 \;,\\
&& \nonumber \\
&& \varphi_0'=V_0' \;.
\end{eqnarray}
For the action~(\ref{2.1}) these become
\begin{eqnarray}
&& H_0^2=\frac{\kappa^2 V_0}{12\left( 1+f_0 \right)} \;,
\label{questa}\\
&&\nonumber \\
&& V_0'=\frac{1+f_0'}{1+f_0} \, V_0 \;,
\end{eqnarray}
where $R_0=12H_0^2$ for de Sitter space. Upon use of
Eq.~(\ref{questa}), the stability
condition~(\ref{stability}) in this theory becomes
\be
\frac{ \left( V_0 f_0'' -2V_0 '' \right)\left( 1+f_0
\right)+V_0 ( f_0')^2}{2\kappa^2 \omega_0
\left( 1+ f_0 \right) +3(f_0')^2} \leq 0 \;.
\ee
These stability conditions are important in the study of
the future evolution
of the late-time accelerated era.

\section{Late time acceleration and inflation with several
scalar fields}

In this section we begin by considering a model with two scalar
fields minimally coupled to gravity (see \cite{UPIAC, LTC, MG}).
Such models are used, for example, in reheating scenarios after
inflation.

An additional degree of freedom appears in this case, so that
for a given solution we may
choose different conditions
on the scalar fields, as shown below. It is possible to restrict
these conditions by studying the perturbative regime for each solution. The action
we consider is
\begin{equation}
S=\int\sqrt{-g}\left[\frac{R}{2\kappa^{2}}- \omega(\phi)
\partial_{\mu}\phi\partial^{\mu}\phi-\sigma(\chi)\partial_{\mu}\chi
\partial^{\mu}\chi-V(\phi,\chi)\right]\ , \label{3.1}
\end{equation}
where $ \omega(\phi)$ and $\sigma(\chi)$ are the kinetic terms,
which depend
on the fields $\phi$ and $\chi$, respectively. We again assume a flat
FRW metric.
The Friedmann equations are written as
\be
H^{2}=\frac{\kappa^{2}}{3}\left[\frac{1}{2}
\omega(\phi)\dot{\phi}^{2}
+ \frac{1}{2}\sigma(\chi)\dot{\chi}^{2}+V(\phi,\chi)\right]\ ,\quad
\dot{H}=-\frac{\kappa^{2}}{2}\left[ \omega(\phi)\dot{\phi}^{2}
+\sigma(\chi)\dot{\chi}^{2}\right]\ . \label{3.3}
\ee
By means of a convenient transformation, we can always redefine
the scalar fields so that we can write $\phi=\chi=t$. The
scalar field equations are given by
\be
\omega(\phi)\ddot{\phi}+\frac{1}{2} \omega'(\phi)\dot{\phi}^{2}
+3H \omega(\phi)\dot{\phi}+\frac{\partial V(\phi,\chi)}{\partial
\phi}=0\ , \quad
\sigma(\chi)\ddot{\chi}+\frac{1}{2}\sigma'(\chi)\dot{\chi}^{2}
+3H\sigma(\chi)\dot{\chi}+\frac{\partial V(\phi,\chi)}{\partial \chi}=0\ . \label{3.4}
\ee
Then, for a given solution $H(t)=f(t)$, and combining
the first Friedmann equation with each scalar field equation, respectively, we find
\be
\omega(\phi)=-\frac{2}{\kappa^{2}}\frac{\partial
f(\phi,\chi)}{\partial \phi}\ , \quad
\sigma(\chi)=-\frac{2}{\kappa^{2}}\frac{\partial f(\phi,\chi)}{\partial \chi}\ , \label{3.5}
\ee
where the function $f(\phi,\chi)$ carries down to $f(t,t)\equiv f(t)$, and is defined
as
\begin{equation}
f(\phi,\chi)=-\frac{\kappa^{2}}{2}\left[\int \omega(\phi)d\phi
+\int\sigma(\chi)d\chi\right]\ .\label{3.6}
\end{equation}
The scalar potential can be expressed as
\begin{equation}
V(\phi,\chi)=\frac{1}{\kappa^{2}}\left[3f(\phi,\chi)^{2}
+\frac{\partial f(\phi,\chi)}{\partial \phi}
+\frac{\partial f(\phi,\chi)}{\partial \chi}\right]\ ,\label{3.7}
\end{equation}
and the second Friedmann equation reads
\begin{equation}
 -\frac{2}{\kappa^{2}}f'(t)= \omega(t)+\sigma(t). \label{3.8}
\end{equation}
Then, the kinetic functions may be chosen to be
\be
\omega(\phi)=-\frac{2}{\kappa^{2}}\left[f'(\phi)+g(\phi)\right]\ ,\quad
\sigma(\phi)=\frac{2}{\kappa^{2}}g(\chi)\ ,\label{3.9}
\end{equation}
where $g$ is an arbitrary function. Hence, the scalar
field potential is
finally obtained as
\begin{equation}
V(\phi,\chi)=\frac{1}{\kappa^{2}}\left[3f(\phi,\chi)^{2}
+f'(\phi)+g(\phi)-g(\chi)\right]\ . \label{3.9a}
\end{equation}

\subsection{Example 1}

We can consider again the solution (\ref{1.17})
\begin{equation}
f(t)=\frac{H_{0}}{t_{s}-t}+\frac{H_{1}}{t^{2}}\ . \label{3.10}
\end{equation}
This solution, as already seen in Sec.~II, reproduces unified
inflation and late time acceleration in a scalar field model
with matter given by action (\ref{eq:1}). We may
now understand this solution as
derived from the two-scalar field model (\ref{3.1}), where a
degree of freedom is added so that we can choose various types
of scalar kinetic and potential
terms, as shown below for the solution~(\ref{3.10}). Then,
from Eqs.~(\ref{3.9}) and~(\ref{3.10}), the kinetic terms
follow:
\be
\omega(\phi)=-\frac{2}{\kappa^{2}}\left[\frac{H_{0}}{
(t_{s}-\phi)^{2}} -\frac{2H_{1}}{\phi^{3}}+g(\phi)\right]\ ,
\quad \sigma(\phi)=\frac{2}{\kappa^{2}}g(\chi)\ . \label{3.11}
\ee
The function $f(\phi,\chi)$ is
\begin{equation}
f(\phi,\chi)=\frac{H_{0}}{t_{s}-\phi}+\frac{H_{1}}{\phi^{2}}
+\int d\phi \, g(\phi)-\int d\chi \, g(\chi),\label{3.12}
\end{equation}
while the scalar potential is
\begin{equation}
V(\phi,\chi)=\frac{1}{\kappa^{2}}\left[{3f(\phi,\chi)}^{2}
+\frac{H_{0}}{(t_{s}-\phi)^{2}}-\frac{2H_{1}}{\phi^{3}}+g(\phi)
 -g(\chi)\right]\ . \label{3.14}
\end{equation}
This scalar potential leads to the cosmological solution (\ref{1.17}).
%reproduces the solution~(\ref{3.10}) with an arbitrary function $g(t)$.

It is possible to further restrict $g(t)$ by studying
the stability of the system considered. To this end, we define
the functions
\begin{equation}
X_{\phi}=\dot{\phi}\ ,\quad
X_{\chi}=\dot{\chi}\ ,\quad
Y=\frac{f(\phi,\chi)}{H}\ . \label{3.15}
\end{equation}
Then, the Friedmann and scalar field equations can be written as
\bea
&& \frac{dX_{\phi}}{dN}=-\frac{1}{2H}\frac{
\omega'(\phi)}{ \omega(\phi)}
\left(X_{\phi}^{2}-1\right)-3\left(X_{\phi}-Y\right)\ , \\
&&\nonumber \\
&& \frac{dX_{\sigma}}{dN}=-\frac{1}{2H}
\frac{\sigma'(\chi)}{\sigma(\chi)}\left(X_{\chi}^{2}
 -1\right)-3\left(X_{\chi}-Y\right)\ ,\nn
&&\nonumber \\
&& \frac{dY}{dN}=\frac{\kappa^{2}}{2H^{2}}\left[
\omega(\phi)X_{\phi}(YX_{\phi}-1)
+\sigma(\chi)X_{\chi}(YX_{\chi}-1)\right] \ , \label{3.16}
\eea
where $\frac{d}{dN}=\frac{1}{H}\frac{1}{dt}$. At
$X_\phi=X_\chi=Y=1$,
we consider the perturbations
\begin{equation}
X_{\phi}=1+\delta X_{\phi}\ ,\quad X_{\chi}=1+\delta X_{\chi}\ ,\quad
Y=1+\delta Y, \label{3.17}
\end{equation}
then
\begin{equation}
\frac{d}{dN}\left(\begin{array}{c}
\delta X_{\phi}\\
\delta X_{\chi}\\
\delta Y\end{array}\right)=M\left(\begin{array}{c}
\delta X_{\phi}\\
\delta X_{\chi}\\
\delta Y\end{array}\right)\ ,\qquad\qquad
M=\left(\begin{array}{ccc}
 -\frac{ \omega'(\phi)}{H \omega(\phi)}-3 & 0 & 3\\
0 & -\frac{\sigma'(\chi)}{H\sigma(\chi)}-3 & 3\\
\kappa^{2}\frac{ \omega(\phi)}{2H^{2}} &
\kappa^{2}\frac{\sigma(\chi)}{2H^{2}} & \kappa^{2}\frac{
\omega(\phi)+\sigma(\chi)}{2H^{2}}\end{array}\right). \label{3.18}
\end{equation}
The eigenvalue equation is given by
\bea
&& \left(\frac{ \omega'(\phi)}{H \omega(\phi)}+3+\lambda\right)
\left(\frac{\sigma'(\chi)}{H\sigma(\chi)}+3+\lambda\right)
\left(\frac{\kappa^{2}}{2H^{2}}(
\omega(\phi)+\sigma(\phi))-\lambda\right) \nn
&&\nonumber \\
&& +\frac{3\kappa^{2}
\omega(\phi)}{2H^{2}}\left(\frac{\sigma'(\chi)}{H\sigma(\chi)}+3+
\lambda\right)+\frac{3\kappa^{2}\sigma(\chi)}{2H^{2}}
\left(\frac{ \omega'(\phi)}{H \omega(\phi)}+3+\lambda\right)=0\ .
\label{3.18a}
\eea
To avoid divergences in the eigenvalues, we choose the
kinetic functions to satisfy
\begin{equation}
\omega(\phi)\neq 0\ ,\quad
\sigma(\chi)\neq 0\ , \label{3.19}
\end{equation}
hence, the eigenvalues in Eq.~(\ref{3.18a}) are finite.
%The expressions for the kinetic terms (\ref{3.9}) and condition
%(\ref{3.19}) imply
%\begin{equation}
%g(t)>-f'(t)\ . \label{3.20}
%\end{equation}
Summing up, under these conditions, the solution (\ref{3.10}) has no infinite
instability when the transition from the non-phantom to the phantom phase occurs.

As an example, we may choose $g(t)=\alpha/t^{3}$, where
$\alpha$
is a constant that satisfies $\alpha>2H_{1}$. Then, the $f(\phi,\chi)$
function (\ref{3.12}) is given by
\begin{equation}
f(\phi,\chi)=\frac{H_{0}}{t_{s}-\phi}-\frac{(\alpha-2H_{1})}{2\phi^{2}}
+\frac{\alpha}{2\chi^{2}}\ . \label{3.20a}
\end{equation}
As a result, the kinetic terms (\ref{3.11}) are expressed as
\be
\omega(\phi)=-\frac{2}{\kappa^{2}}\left[ \frac{H_{0}}{(t_{s}
 -\phi)^{2}}
 -\frac{2H_{1}}{\phi^{3}}+\frac{\alpha}{\phi^{3}}\right] \ ,\quad
\sigma(\chi)=\frac{2}{\kappa^{2}}\frac{\alpha}{\chi^{3}}\ ,
\label{3.21}
\ee
and the potential reads
\begin{equation}
V(\phi,\chi)=\frac{1}{\kappa^{2}}\left[ 3f(\phi,\chi)^{2}
+\frac{H_{0}}{(t_{s}-\phi)^{2}}
+\frac{\alpha-2H_{1}}{\phi^{3}}-\frac{\alpha}{\chi^{3}}\right] \ .
\label{3.22}
\end{equation}
This potential reproduces the solution above, which unifies inflation
and late time acceleration in the context of scalar-tensor
theories,
involving two scalar fields. Notice that the extra degree of
freedom gives
the possibility to select a different kinetic and scalar
potential in such a manner that we get the same solution.

In the case in which the condition (\ref{3.19}) is not imposed,
the kinetic terms~(\ref{3.9}) may have zeros for $0<t<t_{s}$, so
that the perturbation analysis performed above ceases to be
valid
because some of the eigenvalues
could diverge.

\subsection{General case: $n$ scalar fields}

As a generalization of the action~(\ref{3.1}), we now
consider the corresponding one for $n$
scalar fields,
\begin{equation}
S=\int d^{4}x\sqrt{-g}
\left[\frac{1}{2\kappa^{2}}R-\frac{1}{2}\sum_{i=1}^{n}
\omega_{i}(\phi_{i})\partial_{\mu}\phi_{i}\partial^{\mu}\phi_{i}
 - V(\phi_{1},\phi_{2},\cdots,\phi_{n})\right]\ .
\label{3.23}
\end{equation}
The associated Friedmann equations are
\be
H^{2}=\frac{\kappa^{2}}{3}\left[\sum_{i=1}^n
\frac{1}{2}\omega_{i}
(\phi_{i})\dot{\phi_{i}}^{2}+V(\phi_{1},\cdots ,\phi_{n})\right]\ ,\quad
\dot{H}=-\frac{\kappa^{2}}{2}\left[\sum_{i=1}^n
\omega_{i}(\phi_{i})\dot{\phi_{i}}^{2}\right]\ .
\label{3.24}
\ee
We can proceed analogously to the case of two scalar fields
so that the kinetic terms are written as
\begin{equation}
\sum_{i=1}^n \omega_{i}(t)=-\frac{2}{\kappa^{2}}f'(t)\ ,
\label{3.25}
\end{equation}
hence,
\be
\omega_{i}(\phi_{i})=
 -\frac{2}{\kappa^{2}}\frac{df(\phi_{1},\cdots ,
\phi_{n})}{d\phi_{i}}\ ,\quad
V(\phi_{1},\cdots,\phi_{n})=\frac{1}{\kappa^{2}}\left[3f(\phi_{1},\cdots , \phi_{n})^{2}
+\sum_{i=1}^n \frac{df(\phi_{1},\cdots ,
\phi_{n})}{d\phi_{i}}\right]\ , \label{3.26}
\ee
where $f(t,t,\cdots,t)\equiv f(t)$. Then the following solution is found
\be
\phi_{i}=t\ , \quad H(t)=f(t)\ . \label{3.27}
\ee
 From (\ref{3.25}) we can choose, as done above, the kinetic
terms to be
\be
\omega_{1}(\phi_{1})=
 -\frac{2}{\kappa^{2}}\left[f'(\phi_{1})+g_{2}(\phi_{1})
+\cdots +g_{n}(\phi_{1})\right]\ ,\quad
\omega_{2}(\phi_{2})=\frac{2}{\kappa^{2}}g_{2}(\phi_{2})\ ,\cdots ,\
\omega_{n}(\phi_{n})=\frac{2}{\kappa^2}g_{n}(\phi_{n})\ .
\label{3.28}
\ee
Then, there are $n-1$ arbitrary functions that reproduce the
solution~(\ref{3.27}) so reconstruction may be successfully done. They
could be chosen so that dark
matter is also represented by some of the scalar fields appearing in
the action~(\ref{3.23}).

\section{Reconstruction of inflation and cosmic acceleration from
two-scalar theory}

In the present section, inflation and cosmic acceleration are
reconstructed
separately, by means of a two scalar field model that reproduces
some of the cosmological constraints at each epoch. We explore
an inflationary model in which  the scalar
potential, given for a pair of scalar fields, exhibits an extra
degree of freedom and can be chosen in such way that slow-roll
conditions are satisfied. Also, the cosmic acceleration is
reproduced with a pair of scalar fields plus an ordinary matter
term, in which the values of the observed
cosmological density parameter ($\Omega _{\rm DE}\simeq 0.7$)
and of the EoS parameter ($w_{\rm DE}\simeq -1$) are reproduced
in a quite natural way. For the case of a single scalar, the
reconstruction for similarly distant epochs was given
in Ref.~\cite{Neupane} (for other reconstruction versions,
see also \cite{ohta}).

\subsection{Inflation}

In the previous sections, models describing inflation and
late-time accelerated expansion have been constructed by using
certain convenient scalar-tensor theories. In this
section, we present an inflationary model with two scalar fields,
which can be constructed in such a way that the inflationary
conditions are carefully accounted for. For this purpose, we use
some of the techniques given in the previous
section. The action during the inflationary epoch is written as
\begin{equation}
S=\int \sqrt{-g}\left[ \frac{R}{2\kappa
^{2}}-\frac{1}{2}\partial_{\mu}
\phi \partial ^{\mu }\phi -\frac{1}{2}\partial_{\mu }\chi
\partial^{\mu}\chi -V(\phi ,\chi )\right] \ .
\label{5.1}
\end{equation}
We will show that a general solution can be constructed, where
the scalar field  potential is not completely specified because
of the extra degree of freedom  represented by the second scalar
field added, in a  way similar to the situation occurring in
the previous section. Considering a spatially flat FRW metric,
the Friedmann and scalar field equations are obtained by using
the Einstein equations and varying
the action with respect to both scalar fields:
\begin{equation}
H^{2}=\frac{\kappa ^{2}}{3}\left[
\frac{1}{2}\dot{\phi}^{2}+\frac{1}{2}
\dot{\chi}^{2}+V(\phi ,\chi )\right]\ , \quad \ddot{\phi}+3H\dot{\phi}
+\frac{\partial V(\phi,\chi )}{\partial \phi }=0\ , \quad
\ddot{\chi}+3H\dot{\chi}
+\frac{ \partial V(\phi ,\chi )}{ \partial \chi }=0\ .
\label{5.2}
\end{equation}
We assume that the slow-roll conditions are satisfied, that is,
$\ddot\phi \ll 3H\dot\phi$ $\left(\ddot\chi \ll 3H
\dot\chi\right)$, and ${\dot\phi }^{2} \ll V(\phi ,\chi )$
$\left(\dot\chi^{2} \ll V(\phi ,\chi )\right)$,
in order for inflation to occur. Then, Eqs.~(\ref{5.2}) take the
form
\begin{equation}
H^{2}\approx \frac{\kappa ^{2}}{3}V(\phi ,\chi )\ , \quad
3H\dot{\phi}
+\frac{\partial V(\phi,\chi )}{\partial \phi }\approx 0\ , \quad
3H\dot{\chi}+\frac{\partial V(\phi ,\chi )}{\partial \chi
}\approx 0
\label{5.3a}
\end{equation}
and the slow-roll conditions read
\be
\frac{1}{3\kappa ^{2}}\frac{V_{,i}V^{,i}}{V^{2}} \ll 1\ ,\quad
\frac{1}{3\kappa ^{2}}\frac{\sqrt{V_{,ij}V^{,ij}}}{V} \ll 1 \ .
\label{5.4}
\ee
Here,  $V_{,i}$ denotes the partial derivative of $V$ with
respect to
one of the scalar fields ($i=\phi ,\chi $). As done
previously, a scalar potential $ V(\phi ,\chi )$ can be
constructed, although in this case the conditions for inflation
need to be taken into account. From (\ref{5.3a}), the potential is
given by
\begin{equation}
V(\phi ,\chi )=\frac{3}{\kappa ^{2}}\, H^{2}(\phi ,\chi ) \ .
\label{5.5}
\end{equation}
We can choose this potential such that
\begin{equation}
V(\phi ,\chi )=\frac{3}{\kappa ^{2}}\left[ f^{2}(\phi ,\chi )
+g_{1}(\phi)-g_{2}(\chi )\right] \ .
\label{5.6}
\end{equation}
The three components $f$, $g_{1}$,  and $g_{2}$ are arbitrary
functions, and $g(N)=g_{1}(N)=g_{2}(N)$ where, for convenience,
we use the number of e-folds $N \equiv \ln \frac{a(t)}{a_{i}}$
instead of the cosmic time, and $a_{i}$ denotes the
initial value of the scale factor before inflation. Then, the following solution
is found:
\begin{equation}
H(N)=f(N) \ .
\label{5.7}
\end{equation}
Hence, Eqs.~(\ref{5.3a}) may be expressed as a set
of differential equations with the number of e-folds as
independent variable,
\begin{equation}
3 f^2(N) \frac{d\phi }{d N}+\frac{\partial V(\phi ,\chi
)}{\partial \phi
}\approx 0\ ,\quad 3 f^2(N)
\frac{d\chi }{dN}+\frac{\partial V(\phi ,\chi )}{\partial \chi
}\approx 0 \ .
\label{5.8}
\end{equation}

To illustrate this construction, let us use a simple example.
The  following scalar potential, as a function of the number of
e-folds $N$, is
considered:
\begin{equation}
V(\phi ,\chi ) =\frac{3}{\kappa ^{2}}\left[ H_{0}^{2}N^{2\alpha
}\right] \ ,
\label{5.9}
\end{equation}%
where $\alpha $ and $H_{0}$ are free parameters. By specifying
the arbitrary function $g(N)$, one can find a solution for the
scalar fields.  Let us choose, for the sake of simplicity,
$g(N)=g_{0}N^{2\alpha }$, where $g_{0}$ is a constant,
and $f(\phi ,\chi )=f(\phi )$ ({\em i.e.}, as a function of the
scalar field
$\phi$ only). Then, using Eqs.~(\ref{5.8}), the solutions for
the scalar fields are found to be
\be
\phi (N)=\phi _{0}-\frac{1}{\kappa H_{0}}\sqrt{2\alpha (H_{0}^{2}+g_{0})}\
\ln N \ ,\quad
\chi (N)=\chi _{0}-\frac{1}{\kappa H_{0}}\sqrt{2g_{0}\alpha N} \ ,
\ee
and the scalar potential can be written as
\begin{equation}
V(\phi ,\chi )=\frac{3}{\kappa ^{2}}\left[ \left( H_{0}+g_{0}\right) \exp
\left( \kappa \frac{\sqrt{2\alpha }H_{0}}{\sqrt{H_{0}^{2}+g_{0}}}\left(
\phi_{0}-\phi \right) \right) -g_{0}\left( \frac{\kappa ^{2}H_{0}^{2}
(\chi_{0}-\chi )^{2}}{2g_{0}\alpha }\right) ^{2\alpha }\right]\ .
\end{equation}
We are now able to impose the slow-roll conditions  by
evaluating the slow-roll parameters
\bea
\frac{1}{3\kappa ^{2}}\frac{V_{,i}V^{,i}}{V^{2}} &=&
\frac{2\alpha}{3H_{0}^{2}}\left(
\frac{ (H_{0}+g_{0})^{2}}{H_{0}^{2}+g_{0}}
+\frac{4g_{0}}{N}\right) \ll 1 \ ,\nn
&&\nonumber \\
\frac{1}{3\kappa ^{2}}\frac{\sqrt{V_{,ij}V^{,ij}}}{V} &=&
\frac{2}{3}\alpha ^{2}
\sqrt{\frac{(H_{0}+g_{0})^{2}}{(H_{0}^{2}
+g_{0})^{2}}+\frac{16g_{0}(4\alpha
 -1)^{2}}{4g_{0}^{2}\alpha ^{2}}\frac{1}{N^{2}}} \ll 1\ .
\eea
Hence, we may choose conveniently the free parameters so that
the slow-roll conditions are satisfied, and therefore, inflation
takes place.
 From these expressions we see that the desired conditions will be obtained,
in particular, when $\alpha$ is sufficiently small and/or $H_0$ and $N$ are
large enough. All these regimes help to fulfill the slow-roll
conditions,
in a quite natural way.

\subsection{Cosmic acceleration with a pair of scalar fields}

It is quite reasonable, and rather aesthetic,  to think that the
cosmic acceleration could be driven by the same mechanism as
inflation. To this purpose, we apply the same model with two
scalar fields, with the aim of reproducing late-time
acceleration in a universe filled with a fluid with EoS
$p_{m}=w_{m}\rho _{m}$. The free parameters given by the model
could be adjusted to fit the observational data, as shown below.
We begin with the action representing this model,
\begin{equation}
S=\int d^{4}x\sqrt{-g}\left[ \frac{R}{2\kappa
^{2}}-\frac{1}{2}(\partial \phi )^{2}-\frac{1}{2}(\partial \chi
)^{2}-V(\phi ,\chi )+L_{m}\right] \ .
\label{52.1}
\end{equation}
By assuming a spatially flat FRW metric, one obtains the
Friedmann equations
\begin{equation}
H^{2}=\frac{\kappa ^{2}}{3}\left[
\frac{1}{2}(\dot{\phi})^{2}+\frac{1}{2}(
\dot{\chi})^{2}+V(\phi ,\chi )+\rho _{m}\right] \ ,\quad \dot{H}
=-\frac{\kappa }{2}\left[ \frac{1}{2}\dot{\phi}^{2}+\frac{1}{2}\dot{\chi}^{2}
+V(\phi,\chi )+\rho _{m}\right] \ . \label{52.2}
\end{equation}
Variation of the action~(\ref{52.1}) yields the scalar field
equations
\begin{equation}
\ddot{\phi} +3H\dot{\phi}+V_{,\phi }=0\ , \quad
\ddot{\chi}+3H\dot{\chi}
+V_{,\chi}=0\ .
\label{52.3}
\end{equation}
This set of independent equations may be supplemented by a
fifth one, the conservation of matter-energy density,
$\rho _{m}$,
\begin{equation}
\dot{\rho _{m}}+3H\rho _{m}(1+ w_{m})=0\ .
\label{52.4}
\end{equation}
As done in~\cite{Neupane}, we perform the substitutions
\begin{equation}
\Omega _{m}=\frac{\rho _{m}}{3H^{2}/\kappa ^{2}}\ ,\quad
\Omega _{sf}=
\frac{\frac{1}{2}(\dot{\phi})^{2}+\frac{1}{2}(\dot{\chi})^{2}
+V(\phi ,\chi )}{3H^{2}/\kappa ^{2}}\ ,\quad
\epsilon =\frac{\dot{H}}{H^{2}}\ ,\quad
w_{sf} \equiv \frac{p_{sf}}{\rho
_{sf}}=\frac{\frac{1}{2}(\dot{\phi})^{2}+\frac{1}{2}
(\dot{\chi})^{2}-V(\phi ,\chi )}{\frac{1}{2}(\dot{\phi})^{2}
+\frac{1}{2}(\dot{\chi})^{2}+V(\phi ,\chi )}\ .
\label{52.5}
\end{equation}
For convenience, we consider both scalar fields together under
the subscript $sf$, so that the corresponding density parameter
is $ \Omega _{sf}=\Omega _{\phi}+\Omega _{\chi }$.
Then, using the transformations~(\ref{52.5}), the
Friedmann equations and the scalar equation read
\be
\Omega _{m}+\Omega _{sf} = 1 \ ,\quad
2\epsilon +3(1+ w_{sf})\Omega _{sf}+3(1+ w_{m})\Omega _{m} = 0\ ,\quad
\Omega _{sf}^{\prime }+2\epsilon \Omega _{sf}+3\Omega _{sf}(1+ w_{sf})
=0\ .
\label{52.6}
\ee
Here, the prime denotes differentiation with respect to  the
number of e-folds $N \equiv \ln \frac{a(t)}{const}$. We can now
combine Eqs.~(\ref{52.5}) to write the EoS parameter for
the scalar fields as a function of $\Omega_{sf}$ and of the time
derivative of the scalar fields, namely
\begin{equation}
w_{sf}=\frac{\kappa ^{2}(\phi ^{\prime 2}+\chi ^{\prime 2})
 -3\Omega_{sf}}{3\Omega _{sf}}\ .
\label{52.7}
\end{equation}
Hence, it is possible to find an analytic solution for the equations
(\ref{52.6}), for a given evolution of the scalar fields, as will be seen
below. Before doing that, it is useful to write the effective EoS parameter,
which is given by
\begin{equation}
w_{\rm eff}=-1-\frac{2\epsilon }{3}\ ,\quad \rho_{\rm eff}=\rho _{m}
+\rho_{sf}\ ,\quad p_{\rm eff}=p_{m}+p_{sf} \;,
\label{52.8}
\end{equation}
and the deceleration parameter
\begin{equation}
q=-\frac{\ddot{a}}{aH^{2}}=-1-\epsilon \ . \label{52.9}
\end{equation}%
As usual, for $q>0$ the universe is in a decelerated phase, while $q<0$
denotes an accelerated epoch, such that for $w_{\rm eff}<-1/3$ the
expansion is accelerated. To solve the equations, we consider a
universe
which, at present, is filled with a pressureless component
($w_{m}=0$) representing ordinary matter, and  two scalar
fields which  represent a dynamical dark energy and a dark
matter
species. To show this, we make the following
assumption on the evolution of the scalar fields, which are given as
functions of $N$:
\begin{equation}
\phi (N)=\phi _{0} +\frac{\alpha }{\kappa ^{2}}N\ ,\quad \chi
(N)=\chi _{0}
+ \frac{\beta }{\kappa ^{2}}N\ .
\label{52.10}
\end{equation}
Then, Eqs.~(\ref{52.6}) can be solved,  and the scalar
field density parameter takes the form
\be
\Omega _{sf} =\Omega _{\phi }+\Omega _{\chi }
=1-\frac{\lambda }{k\mathrm{e}^{\lambda N}+3} \ ,
\label{52.11}
\ee
where $k$ is an integration constant and $\lambda=  3-(\alpha^2
+ \beta^2)$. It is possible to introduce an arbitrary function
$g(N)$, to express the energy
density parameter for each scalar field in the following way:
\be
 \Omega_{\phi } = 1-\frac{\lambda }{k\mathrm{e}^{\lambda
N}+3}-g(N) \ ,\quad
\Omega _{\chi } = g(N) \ .
\label{52.12}
\ee
The function $g(N)$ may be chosen in such a way that the
scalar field $\chi$ represents
a cold dark matter contribution at present ($w_{\chi}\simeq 0$),
and the scalar field $\phi$ represents the dark energy
responsible for
the accelerated expansion of our universe. On the other hand,
using Eqs.~(\ref{52.6}), $\epsilon=\frac{H'}{H}$ is obtained as
\be
\epsilon=-\frac{3}{2}\left\{ 1-\frac{k\lambda \left( k\e^{\lambda N}
+\alpha^2\beta^2\right) }{\left[ (\alpha^2+\beta^2)\e^{-\lambda N}
+3k\right]\left(k\e^{\lambda N} +3 \right) }\right\} \ .
\label{52.13}
\ee
 Then, it is possible to calculate the effective parameter of
EoS given
by Eq.~(\ref{5.8})
\be
w_{\rm eff}=-1-\frac{2}{3}\epsilon
= -\frac{k\lambda \left( k\e^{\lambda N} +\alpha^2\beta^2\right) }
{\left[ (\alpha^2+\beta^2)\e^{-\lambda N}+3k\right]\left(k\e^{\lambda N} +3 \right) }\ .
\label{52.14}
\ee
We have four free parameters ($N$, $k$, $\alpha$, $\beta$) that may
 be adjusted to fit  the constraints derived from
observations. With this purpose, we use the observational input
$\Omega_{m}\simeq 0.03$, referred to baryonic matter, and normalize the number
of e-folds $N$, taking $N=0$ at present, then the integration constant $k$ may be written
as a function of $\alpha$ and $\beta$ as
\be
\Omega_m (N=0)=0.03\ \to\ k=\frac{2.01-(\alpha^2 + \beta^2)}{0.03}\ .
\label{52.15}
\ee
The free parameter $\beta$ may be fixed in such a way that the scalar field $\chi$
represents cold dark matter at present, {\em i.e.},
$w_{\chi}\simeq 0$
and $\Omega_{\chi}\simeq0.27$, and its EoS parameter is written
as
\be
w_{\chi}=\frac{\kappa^{2}{\chi'}^{2}
 -3\Omega_{\chi}}{3\Omega_{\chi}}=\frac{\beta^2 -3g(N)}{3g(N)} \ .
\label{52.16}
\ee
For convenience, we choose $g(N)=\frac{\beta^2}{3}\e^{-N}$, then the energy density
and EoS parameter are given by
\be
w_{\chi}=\frac{1-\e^{-N}}{\e^{-N}}\ ,\quad
\Omega_{\chi}=\frac{\beta^2}{3}\e^{-N}\ .
\label{52.17}
\ee
Hence, at present ($N=0$), the expressions~(\ref{52.17}) can be
compared with
the observational values and the $\beta$ parameter is given by
\be
w_{\chi}(N=0)=0\ ,\quad
\Omega_{\chi}(N=0)\simeq 0.27\ \to\ \beta^2=0.81\ .
\label{5.18}
\ee
Finally, the energy density of $\phi$ expressed by
Eq.~(\ref{52.12}) takes the form
\be
\Omega _{\phi } =1-\frac{\lambda }{k\mathrm{e}^{\lambda N}+3}-\frac{\beta^2}{3}\e^{-N}\ .
\ee
The value for $\alpha$ could be taken so that
$\Omega_{\phi}\simeq 0.7$
and $w_{\phi}\simeq -1$ at present. Hence, it has been shown that
cosmic acceleration can be reproduced with a pair of scalar
fields,
where due to the presence of the extra scalar that can
be identified with the
dark matter component.

It is interesting  to point out that one can unify these
realistic descriptions of the inflationary and late-time
acceleration eras within a single theory. However, the
corresponding potential looks quite complicated. The easiest
way would be to use step ($\theta$-function) potentials in order
to  unify the whole description in the easiest way (as was
pioneered in \cite{cogn08a}).

We may also construct a model unifying early  universe inflation
and the present accelerated expansion era. To
this end we can choose $f(\phi)$ in (\ref{uf0}), which gives the Hubble
parameter~(\ref{uf1}). Then, using (\ref{3.9}), one can define
$\omega(\phi)$ and $\sigma(\chi)$ with the help of an arbitrary
function $g$. After defining then $f(\phi,\chi)$
with Eq.~(\ref{3.6}), we can construct the potential
$V(\phi,\chi)$ using Eq.~(\ref{3.7}). Finally, we obtain the
two scalar-tensor theory (\ref{3.1})
reproducing the Hubble rate (\ref{uf1}), which describes both
inflation and the accelerated
expansion.

\section{Conclusions}

Modeling both the early inflation and late-time
acceleration epochs within the context of a single field theory has,
undoubtedly, much aesthetic appeal and seems a worthy goal,
which we have attempted here.  To summarize, we have
developed, step by step, the reconstruction program for
the expansion history of the universe, by using a single or
multiple
(canonical and/or phantom) scalar fields. Already in the case of a
single scalar, we  have  presented a number of examples
which prove that it is actually possible to unify early-time
inflation (at very high redshift)  with late-time acceleration (at
low redshift). The reconstruction
technique has then been  generalized to the case of
a scalar field non-minimally coupled to the Ricci curvature, and
to non-minimal (Brans-Dicke-type) scalars.
Again, various explicit examples of unification of early-time
inflation and late-time acceleration have been presented in those
formulations.
Due to the special role of de Sitter space, which often appears as an
attractor in the  inflationary
epoch, as well as in the present cosmic acceleration era,
special attention was paid to this specific space-time. Conformal
transformations to the Einstein frame  and stability conditions
for the de Sitter space were discussed.

Moreover, the case of several minimally coupled  scalar fields
has  been considered for the description of the realistic
evolution of  the Hubble parameter, and we have  shown that it
is  qualitatively easier to achieve a realistic
unification of late and early epochs in a model of this kind, in such
a way as to satisfy the cosmological
bounds coming from the observational data. This is due to the
arbitrariness in the choice of the scalar potential and
the scalar kinetic factor in the description of a  universe with
a given scale factor $a(t)$. Using the
freedom of choosing these scalar functions, one can constrain
the theory in an observationally acceptable manner. Specifically,
slow-roll conditions and stability conditions may be satisfied
in different ways for different scalar functions,  while the
scale factor remains the same. This can be  used also to obtain
the correct structure of perturbations, {\em etc}.

As a mater of fact, many questions remain to be discussed in greater
detail, a more realistic matter content should be taken into
account, and the universe expansion
history be described in a  more precise and  detailed manner.
After all, we live in an era of increasingly more precise
cosmological tests. Anyhow, the
unified effective description of the cosmic expansion history
presented here seems quite promising. Using it in
more  realistic contexts---in which, of
course, technical details become necessarily more complicated---appears
to be quite possible.

\begin{acknowledgments}
EE, SDO and DS were supported, in part, by MEC (Spain), project
FIS2006-02842
and by AGAUR (Gene\-ra\-litat de Ca\-ta\-lu\-nya), 
contract 2005SGR-00790 and grant 2007BE-1003.
SDO has been  supported also by MEC (Spain) projects
PIE2007-50/023 and FIS2005-01181 and RFBR grant 06-01-00609.
VF acknowledges financial support from the Natural Sciences
and Engineering Research Council of Canada (NSERC).
The research of SN has been supported, in part, by the Ministry
of Education, Science, Sports and Culture of Japan under grant
no.18549001 and 21st Century COE Program of Nagoya University
provided by the Japan Society for the Promotion of Science
(15COEG01).

\end{acknowledgments}

\end{document}